\documentstyle[aps]{revtex}

\begin{document}

\begin{titlepage}
 {$\mbox{}$}
\vskip 2cm
\centerline{\large\bf CHIRAL AND SPIN ORDER IN THE
2D $\pm J$ XY SPIN GLASS : }
\vskip 0,5cm
\centerline{\large\bf DOMAIN WALL SCALING ANALYSIS}
\vskip 4cm
\renewcommand{\thefootnote}{\fnsymbol{footnote}}
\centerline{\large M. Ney-Nifle and H.J. Hilhorst}
\vskip 0,3cm
\centerline{\large Laboratoire de Physique Th\'eorique
et Hautes Energies
\footnote[1]{\noindent \normalsize Laboratoire associ\'e
au Centre National de la Recherche Scientifique}}
\vskip 0,2cm
\centerline{\large B\^atiment 211, Universit\'e de
Paris-Sud, 91405 Orsay C\'edex}
\vskip 5cm

{\centerline \today}

\vskip 5cm

\vskip 0,3cm

\end{titlepage}

\begin{abstract}

This is an analytic study of the two-dimensional XY
spin glass with $\pm J$ disorder. The Hamiltonian has
a continuous spin symmetry and a discrete chiral symmetry,
and therefore possesses, potentially, two
different order parameters and correlation lengths.
The cost of breaking
the symmetries is probed by comparing the ground state
energy under
periodic (P) boundary conditions with the one under
antiperiodic (AP) and
under reflecting (R) boundary conditions. Two energy
differences
(``domain wall energies'') appear, $\Delta E^{AP}$ and
$\Delta E^R$,
whose scaling behavior with system size is nontrivially
related to the
correlation length exponents.

For a specific distribution of the $\pm J$ disorder
 we show that the
chiral and spin correlation lengths diverge with the
same exponent as $T
\downarrow 0$. The common exponent has a common cause,
viz. the low reversal energy of
domains of chiral variables. For general disorder we
give a
heuristic argumentation in terms of droplet excitations
that leads to
spin ordering on a longer, or equal, scale than chiral
ordering. These
results are in contrast with interpretations of
Monte Carlo
simulations.

\end{abstract}

\section{Introduction}
Due to the
rotational symmetry of the XY  model, any of its
ground states is
necessarily part of a continuum of ground states
 related by global spin
rotations. It was first pointed out by Villain [1]
 that the ground state
of an XY spin glass with random $\pm J$ interactions
 has, in addition, a
twofold degeneracy. The two continua can be deduced
 from one another by
a global spin reflection with respect to an arbitrary
 axis. They are
characterized by opposite ``chiralities'', that is,
 by an opposite sense
of rotation of the spins as one moves around a
plaquette of
the lattice. At finite temperature one may have
 domains
belonging to different ground states (``chiral
excitations''), so that each plaquette has to be
characterized by its own chiral variable.

A number of recent theoretical papers have focused on
the r\^ole of these chiralities [2-6]. The motivation
 for this interest is the question what the lower critical
dimension $d_\ell$ is of the XY spin glass, and by which
mechanism this system orders just above $d_\ell$. Ozeki
and Nishimori [7-9] have developed arguments for the bound
$d_\ell \geq 4$ (see also Schwartz and Young [10]), which
is supported by substantial numerical evidence (see
references in [9]). However, all these authors consider
the conventional Edwards-Anderson order parameter
associated with the XY spins. Villain's [1] discovery
 of a discrete
symmetry therefore naturally led to the idea, due to
Kawamura and
Tanemura [2, 3], that in fact $d_\ell$ might be less
than four. We
recall briefly how this idea has been investigated in
 recent years.

One can define two correlation lengths
$\xi_c$ and $\xi_s$, associated with the chiral variables
$q_{\mbox{\boldmath $r$}}$ (where ${\mbox{\boldmath $r$}}$
denotes the center of a plaquette) and the
spin variables ${\mbox{\boldmath $S$}}_{\mbox{\boldmath $i$}}$
 (where ${\mbox{\boldmath $i$}}$ denotes a lattice
site), respectively :
$$
\overline{<q_{\mbox{\boldmath $r$}} q_{{\mbox{\boldmath $r$}}
+{\mbox{\boldmath $R$}}}>^2} \sim e^{-R/\xi_c} \qquad
\hbox{and} \qquad
\overline{<{\mbox{\boldmath $S$}}_{\mbox{\boldmath $i$}}
\cdot {\mbox{\boldmath $S$}}_{{\mbox{\boldmath $i$}} +
{\mbox{\boldmath $R$}} }>^2} \sim {e}^{- R/\xi_s} \qquad
(R\rightarrow \infty) \ \ \ . \eqno(1.1) $$
Here $<\cdots>$ denotes the thermal and $\overline{\cdots}$
the disorder average. For the low
temperature behavior of these correlation lengths one
expects for $d < d_{\ell}$ :
$$
\xi_c(T) \sim T^{-\nu_c} \qquad \hbox{and} \qquad
\xi_s(T) \sim
T^{-\nu_s} \qquad (T \rightarrow 0) \ \ \ . \eqno (1.2)
$$
One method to investigate the relation between $\xi_c$
and $\xi_s$ is to determine
the correlation length exponents $\nu_c$ and $\nu_s$
via finite-size
scaling of ``domain wall'' energies. In  simpler cases
where there is a single symmetry and a single correlation
length
with exponent $\nu$, this is done by finding the ground
state energies
for two appropriately chosen distinct boundary conditions.
The
energy difference, $\Delta E$, will then generally scale
with the linear system size $N$ as $\overline{|\Delta E|}
\sim N^{-1/\nu}$.

In the present case, with both a continuous and a discrete
symmetry, the periodic (P) ``reference'' boundary
conditions can be changed in two ways : to antiperiodic
(AP) and to reflecting (R) ones, this latter possibility
having been first suggested by Kawamura and Tanemura [3].
Hence there are two energy differences to be considered,
$\Delta E^{AP}$
and $\Delta E^R$. Across an antiferromagnetic seam two
spins interact with each other's image under reflection
in the origin (of
spin space). Hence AP boundary conditions probe the
spatial rigidity
of spin order, and the spin correlation length exponent
$\nu_s$ is given
by the usual relation
$$
\overline{|\Delta E^{AP}|} \sim N^{-1/\nu_s} \eqno(1.3)
$$
Across a reflecting seam two spins interact with each
other's image
under reflection about a fixed axis (in spin space).
Since the chiral
variables appear only when the ``spin waves'' are
integrated out - and
notwithstanding the obvious link between R boundary
conditions and
chirality - it is not a priori clear exactly how these
boundary
conditions probe the chiral order. In fact the relation
$\overline{|\Delta E^R|} \sim N^{-1/\nu_c}$ that one
might naively expect
is certainly not correct. The reason is that  R boundary
conditions, upon
closer inspection, appear to affect also the spin
wave degrees of
freedom. This was first realized with remarkable intuition
by  Kawamura
and Tanemura [3], who then proceeded and extracted the
chiral correlation
length exponent $\nu_c$ from a heuristic expression
involving both
$\Delta E^{AP}$ and $\Delta E^{R}$.

On the basis of Monte Carlo determinations of the ground
states of two- and three-dimensional systems these authors
conclude that in two
dimensions $\nu_s = 1.2 \pm 0.15$ and $\nu_c = 2.6 \pm 0.3$.
This
implies that the chiral variables have their own length scale
and order
in larger domains than the spin variables. Similar conclusions,
namely
$\nu_s \approx 1$ and $\nu_c \approx 2$, were reached by
Ray and Moore
[5], also on the basis of Monte Carlo simulations and a
finite size
scaling analysis. These studies suggest that as the dimension
$d$ is
increased, $\nu_c^{-1}$ may vanish before $\nu^{-1}_s$
does, so that
just above $d_\ell$ there would be a phase with long
range chiral order
but exponentially decaying spin-spin correlations.
Indeed this is
precisely the scenario that Kawamura and Tanemura [2,3]
find in Monte
Carlo simulations of the two- and three-dimensional XY
spin glass. In recent
work on the two- and three-dimensional Heisenberg spin
glass Kawamura
[11] arrives at fully analogous conclusions.
If these  are accepted, the XY and Heisenberg vector spin
glasses have their lower critical dimensionality between
$d=2$ and $d=3$.

In order to achieve a better understanding of the interplay
between
chiral and spin variables, Ney-Nifle, Hilhorst, and Moore [6]
recently
performed an analytic study of the random $\pm J$ XY spin
glass (in its
Villain [12] formulation) on the one-dimensional ladder
lattice (see
also Morris et al. [13]). Their
conclusion is that the chiral correlation length $\xi_c$
and spin
correlation length $\xi_s$ diverge for $T \downarrow 0$
with the same
exponent $\nu_c = \nu_s$ (of which they determine the
value 1.8999
...). The purpose of this work is
to extend the methods of Ref. [6] to $d=2$ and to confront
our results
with those of Refs. [3] and [5].

Our starting point in Section 2 is the Villain [12]
formulation of the
XY model with random $\pm J$ interactions. This formulation
has the
advantage that the chiral variables can easily be defined
and can be
decoupled from the spin waves. We transform the XY
Hamiltonian to a
Coulomb gas Hamiltonian with charges $q_{\mbox{\boldmath $r$}}$
that
play the r\^ole of
the chiral variables. The transformation has been
known [14-16] since
long, including the fact [1, 17] that the charges take
half-integer values on the frustrated plaquettes and
integer values
on the others. Here we take exactly into account
all finite size effects, essential for the finite size
scaling analysis
that follows. Our  result for the Coulomb gas partition
function in the
case of periodic (P) boundary conditions, and for an
arbitrary
realization of the disorder, is given by Eqs. (2.38).
The effective
Hamiltonian contains, in addition to the Coulomb
interaction, a
coupling term between the total electric dipole moment
and the
boundary conditions. This additional term was obtained
via a different
approach in a very recent paper by Vallat and Beck [18],
and its one-dimensional equivalent had appeared in [6].
In Sections 3 and 4 we go further and discuss how these
equations
are modified in the case of antiperiodic (AP) and reflecting
(R) boundary
conditions, respectively. In the limit $T \rightarrow 0$ the
equations reduce to expressions for the ground state energy,
including
all its finite size corrections, {\it provided} the ground
state itself
is known or can be plausibly guessed.

Since the ground state problem for an arbitrary realization
of the
disorder cannot be solved, we treat in Section 5 a more
restricted
two-parameter subset of realizations, in which the frustrated
plaquettes are placed in a random rectangular array with
infinite
range correlation along the $y$ direction. (Disorder of the
unidirectionally infinite-ranged type was also the first one
considered for the random bond Ising model [19].) We find a
regime of main interest in parameter space in which the XY
spin glass has
a zero-temperature transition with
$$
\nu_s = \nu_c \eqno (1.4)
$$
There is a second regime, with a strongly anisotropic spatial
distribution of the frustrated plaquettes, in which the model has a
low temperature phase with long range chiral correlation and, very
plausibly, power law decay of spin correlations. This is behavior
analogous to that of  fully frustrated two-dimensional XY models
[20-22], on which there exists a large literature.

The result (1.4) in the first regime is different from the scenario
proposed by Kawamura and Tanemura [2, 3] and by Ray and Moore [5].
The
basic mechanism responsible for (1.4) is that the ground state can
accommodate to a changeover from P to AP boundary conditions by
means of
the formation of a chiral domain wall, which is energetically
lower lying
than the continuous spin wave deformation that naturally comes
to mind. On the basis of this example alone we cannot rule
 out the possibility that our conclusion, Eq.
(1.4), is valid only within the restricted class of disorder
realizations.

In order to treat the case of general disorder, we construct in
Section 6 a heuristic
theory based on the same mechanism. It involves the lowest lying
excitations of the standard Coulomb Hamiltonian which are assumed
to be collective charges reversals. We find that
$$
\nu_s \geq \nu_c \ . \eqno (1.5)
$$
This is compatible with (1.4) but
contradicts the results
of Refs.[2,3,5] where the opposite inequality holds.
Furthermore, we show that if Eq. (1.5) would holds
as a strict inequality,
one cannot extract the exponent $\nu_c$ from
$\Delta
E^{R}$ and  $\Delta E^{AP}$. To find $\nu_c$, an appropriate
quantity would then be the energy difference
between the two ground states
of the Coulomb
Hamiltonian with periodic and reflecting boundary conditions,
without any additional terms.

Our conclusion is that there is no evidence that chiral
order extends on a longer length scale than spin order.

\section{Partition function with periodic boundary conditions}

\subsection{The Villain $XY$ model with $\pm \ J$ interactions}

We consider an $XY$ model on a finite square lattice,
periodic in both directions, with sites ${\mbox{\boldmath $i$} } =
(m, n)$, where $m = 1, \cdots, M$ and $n=1, \cdots,
N$. Each site ${\mbox{\boldmath $i$}}$ is occupied by a
two-component unit
vector or ``spin''
${\mbox{\boldmath $S$}}_{\mbox{\boldmath $i$}}$ whose angle
$\phi_{\mbox{\boldmath $i$}}$ with a reference axis takes values in
$(-\pi, \pi]$. In the ferromagnetic Villain model [12] two spins
$\phi_{\mbox{\boldmath $i$}}$ and $\phi_{\mbox{\boldmath $j$}}$
linked by a nearest-neighbor bond
$<{\mbox{\boldmath $i$}}, {\mbox{\boldmath $j$}}>$ have the
Boltzmann weight
$$
e^{-\beta J V(\phi_{\mbox{\boldmath $i$}} -
\phi_{\mbox{\boldmath $j$}})} \equiv
\sum_{n=-\infty}^\infty e^{-\beta J(\phi_{\mbox{\boldmath $i$}} -
\phi_{\mbox{\boldmath $j$}} - 2 \pi n)^2} \ \ \ . \eqno (2.1)
$$
Here $J > 0$ sets the energy scale. The relation (2.1)
implies that the Villain interaction $V(\phi)$
depends on $\beta J$ ;  when $\beta$
becomes large, only one term on the RHS of (2.1) will
dominate, $V(\phi)$ will tend to
$$
V(\phi) = \phi^2 \qquad \hbox{for}\  |\phi| \leq \pi,
\qquad \beta = \infty \eqno (2.2)
$$
and we may interpret $\beta$ as the
inverse temperature $1/k_BT$. For all $\beta$, the sum
on $n$ in (2.1) guarantees the periodicity property
$V(\varphi) = V(\varphi + 2\pi)$ and the symmetry property
$V(\varphi) = V(-\varphi)$. The antiferromagnetic
Villain model has $V(\phi_{\mbox{\boldmath $i$}} -
\phi_{\mbox{\boldmath $j$}}-\pi)$
instead of $V(\phi_{\mbox{\boldmath $i$}} -
\phi_{\mbox{\boldmath $j$}})$.

Here we wish to consider an XY spin glass with
randomly ferro- or antiferromagnetic Villain
interactions, that is, with the Hamiltonian
$$
{\cal H} = J \sum_{<\mbox{\boldmath $i$},
\mbox{\boldmath $j$}>} V(\phi_{\mbox{\boldmath $i$}} -
\phi_{\mbox{\boldmath $j$}} - \pi_{\mbox{\boldmath $ij$}})
\eqno (2.3)
$$
where the sum runs over all nearest-neighbor bonds of
the periodic lattice, and the $\pi_{\mbox{\boldmath $ij$}}$
are quenched random variables such that
$$
\pi_{\mbox{\boldmath $ij$}} = \left  \{{0\atop \pi}\right \}
\ \hbox{with
probability }\  {1\over 2} \eqno (2.4)
$$
The expression for the canonical partition function of
this XY spin glass then is
$$
Z_{M,N} = \int^\pi_{-\pi} \prod_{\mbox{\boldmath $i$}} d
\varphi_{\mbox{\boldmath $i$}}
\sum_{\{n_{\mbox{\boldmath $ij$}}\}} e^{-\beta J
\sum\limits_{< \mbox{\small\boldmath
$i$}, {\mbox{\small\boldmath $j$}}>}
(\varphi_{\mbox{\boldmath $i$}} -
\varphi_{\mbox{\boldmath $j$}} - \pi_{\mbox{\boldmath $ij$}}
- 2\pi n_{\mbox{\boldmath $ij$}})^2} \eqno (2.5)
$$
in which the $n_{\mbox{\boldmath $ij$}}$ may be seen as
additional
dynamical variables, and the argument of the
exponential as a new effective Hamiltonian. We shall
henceforth write
$$
\nu_{\mbox{\boldmath $ij$}} \equiv
n_{\mbox{\boldmath $ij$}}/(2\pi) \ \ \ , \eqno (2.6)
$$
which is integer (half-integer) when the interaction
between $\varphi_{\mbox{\boldmath $i$}}$ and
$\varphi_{\mbox{\boldmath $j$}}$ is
ferromagnetic (antiferromagnetic).

\subsection{Transformation to a Coulomb gas}

\subsubsection{From XY spin glass to a SOS model}

The transformation from an XY model to an SOS model is well-known
[14,15] and has been extended to various types of random XY models.
Here we carefully study the finite size effects, that determine
the ground state energy differences under different boundary
conditions.
In (2.5) we wish to carry out the integrations on the $MN$
variables
$\varphi_{\mbox{\boldmath $i$}}$. To that end we arbitrarily
select a lattice site
${\mbox{\boldmath $i$}}_0$ and transform to the new
variables of integration

$$\varphi_0 = \varphi_{\mbox{\boldmath $i$}_0}
\eqno(2.7\mbox{a})$$

$$\varphi_{\mbox{\boldmath $ij$}} = \varphi_{\mbox{\boldmath $i$}} -
\varphi_{\mbox{\boldmath $j$}}
\ \ \ . \eqno(2.7\mbox{b})
$$
These are $2MN + 1$ in number, and there exist $MN+1$
relations between
them. In order to formulate these, let $\mbox{\boldmath $r$}$,
$\mbox{\boldmath $s$}$, $\cdots$ be
the position vectors of the plaquette centers. We adopt the
convention
that in $<\mbox{\boldmath $i$}, \mbox{\boldmath $j$}>$ the
site $\mbox{\boldmath $j$}$ is to the right of $\mbox{\boldmath
$i$}$ (for a horizontal bond) or above $\mbox{\boldmath $i$}$
(for a vertical bond),
which naturally extends across the periodic boundaries.
We define
furthermore a sum on the bonds $<\mbox{\boldmath $i$},
\mbox{\boldmath $i'$}>$ surrounding a
plaquette $\mbox{\boldmath $r$}$ by the diagram of Fig. 1
\begin{figure}
\setlength{\unitlength}{1cm}
\begin{picture}(19,6)
\thicklines
\put(7,1){\line(0,1){4}}
\put(11,1){\line(0,1){4}}
\put(7,1){\line(1,0){4}}
\put(7,5){\line(1,0){4}}
\put(6.5,0.5){\makebox(0,0){$1$}}
\put(11.5,0.5){\makebox(0,0){$2$}}
\put(11.5,5.5){\makebox(0,0){$3$}}
\put(6.5,5.5){\makebox(0,0){$4$}}
\put(6.5,3){\makebox(0,0){$\varphi_{14}$}}
\put(11.5,3){\makebox(0,0){$\varphi_{23}$}}
\put(9,0.5){\makebox(0,0){$\varphi_{12}$}}
\put(9,5.5){\makebox(0,0){$\varphi_{43}$}}
\put(7.5,3){\makebox(0,0){$-$}}
\put(10.5,3){\makebox(0,0){$+$}}
\put(9,1.5){\makebox(0,0){$+$}}
\put(9,4.5){\makebox(0,0){$-$}}
\put(9,3.3){\makebox(0,0){$r$}}
\put(9,3){\circle*{0.2}}
\put(7.5,3){\circle{0.5}}
\put(10.5,3){\circle{0.5}}
\put(9,1.5){\circle{0.5}}
\put(9,4.5){\circle{0.5}}
\end{picture}

{\footnotesize {\noindent\bf\footnotesize Figure 1:}
{\sl\footnotesize Diagram indicating the sign convention
in the
plaquette sum of Eq. (2.8).} The indices $1, \cdots, 4$
are shorthand for
${\mbox{\boldmath $i$}}_1 , \cdots ,
{\mbox{\boldmath $i$}}_4$.}
\end{figure}
together with the formula

$$\widehat{\sum}^{\mbox{\boldmath $r$}}
\varphi_{\mbox{\boldmath $ii'$}} \equiv
\varphi_{12} + \varphi_{23} - \varphi_{43} -
\varphi_{14} \ \ \ .
\eqno(2.8)$$ \noindent The hat on the summation sign
is a reminder of
the sign convention in the RHS of (2.8). The variables
$\varphi_{\mbox{\boldmath $ij$}}$ of (2.7b) then satisfy
the relations

$$\widehat{\sum}^{\mbox{\boldmath $r$}}
\varphi_{\mbox{\boldmath $ii'$}} = 0 \ \hbox{mod} \
2 \pi \qquad \hbox{for \ all} \ \mbox{\boldmath $r$}
\ \ \ . \eqno(2.9\mbox{a})$$

\noindent These are $MN$ relations of which only $MN - 1$
are independent. The two remaining
relations correspond to loops around the torus and are

$${\sum}^x \ \varphi_{{\mbox{\boldmath $ij$}}} \equiv
\sum_{m=1}^M
\varphi_{(m,n_1), (m+1, n_1)} = 0 \ \hbox{mod} \
2 \pi \eqno(2.9\mbox{b})$$

$${\sum}^y \ \varphi_{{\mbox{\boldmath $ij$}}} \equiv
\sum_{n=1}^N
\varphi_{(m_1, n),(m_1, n+1)} = 0 \ \hbox{mod} \ 2 \pi
\eqno(2.9\mbox{c})$$

\noindent where ${\mbox{\boldmath $i$}}_1 \equiv
(m_1, n_1)$ is another arbitrarily selected lattice
site. Upon
introducing the variables of integration (2.7) in
Eq. (2.5) we must
represent the conditions (2.9) by delta functions
and find
\setcounter{equation}{9}
\begin{equation}
\begin{array}{lcl}
Z_{M, N} &= &2 \pi \int_{- \infty}^{\infty}
\prod_{<{\mbox{\boldmath $i$}}, {\mbox{\boldmath $j$}}>}
d \varphi_{{\mbox{\boldmath $ij$}}}
\prod_{{\mbox{\boldmath $r$}} \not=
{\mbox{\boldmath $r$}}_0} \delta \left (
\widehat{\sum}^{{\mbox{\boldmath $r$}}}
\varphi_{{\mbox{\boldmath $ij$}}} \ \hbox{mod}
\ 2 \pi \right ) \\[0.2cm]
{}&${}$ &\times \delta \left ( {\sum}^x
\varphi_{{\mbox{\boldmath $ij$}}} \ \hbox{mod}
\ 2 \pi \right ) \delta \left ( {\sum}^y
\ \varphi_{{\mbox{\boldmath $ij$}}} \
\hbox{mod} \ 2 \pi \right )\\[0.2cm]
{}&${}$ &\times \exp \left [ - \beta J
\sum\limits_{<{\mbox{\boldmath $ij$}}>}
\left ( \varphi_{{\mbox{\boldmath $ij$}}} -
\pi_{{\mbox{\boldmath $ij$}}} \right )^2 \right ]
\end{array}
\end{equation}

\noindent in which ${\mbox{\boldmath $r$}}_0$ is an
arbitrarily selected plaquette, and where a trivial factor
$2 \pi$ comes from the integration on $\varphi_0$.
The delta functions may be represented by a
Fourier sum with the aid of

$$\delta \left ( x \ \hbox{mod} \ 2 \pi \right ) =
\sum_{\ell = - \infty}^{\infty} \delta (x - 2
\pi \ \ell ) = (2 \pi)^{-1} \sum_{\ell = -
\infty}^{\infty} e^{i \ \ell \ x} \eqno(2.11)$$

\noindent which requires the introduction of a
set $\{ n_{{\mbox{\boldmath $r$}}} \}$ of summation variables
for the plaquettes, and of two summation variables
$n_x$ and $n_y$ for the loops around the
torus. One finds
$$Z_{M,N} = (2 \pi)^{-MN} \sum_{\{n_{{\mbox{\boldmath $r$}}}
\}}{}' \sum_{n_x}
\sum_{n_y} \int_{- \infty}^{\infty}
\prod_{<{\mbox{\boldmath $i$}} , {\mbox{\boldmath $j$}}>} d
\varphi_{{\mbox{\boldmath $ij$}}} \ \exp
\left [ i \sum_{{\mbox{\boldmath $r$}}}
n_{{\mbox{\boldmath $r$}}}
\widehat{\sum}^{{\mbox{\boldmath $r$}}}
\varphi_{{\mbox{\boldmath $ij$}}} \right ]$$
$$\times \  \exp \left [ in_x {\sum}^x
\varphi_{{\mbox{\boldmath $ij$}}} + in_y
{\sum}^y \varphi_{{\mbox{\boldmath $ij$}}}
\right ]$$  $$\times \ \exp \left [
- \beta J \sum_{<{\mbox{\boldmath $i$}} ,
{\mbox{\boldmath $j$}}>} \left (
\varphi_{{\mbox{\boldmath $ij$}}}
- \pi_{{\mbox{\boldmath $ij$}}} \right )^2
\right ]  \eqno(2.12)$$

\noindent where the prime on the summation sign
indicates the restriction to $n_{{\mbox{\boldmath
$r$}}_0} =
0$. The integrations on the $\varphi_{{\mbox{\boldmath
$ij$}}}$ can now be
performed. \par Let the geometric relation between the
pair of lattice
sites $<{\mbox{\boldmath $i$}}, {\mbox{\boldmath $j$}}>$
and the pair of plaquette centers $< {\mbox{\boldmath $r$}}
, {\mbox{\boldmath $s$}}>$ be as in Fig. 2.
\begin{figure}
\setlength{\unitlength}{1cm}
\begin{picture}(19,6)
\thicklines
\put(3,3){\line(1,0){4}}
\put(11,3){\line(1,0){4}}
\put(5,1){\line(0,1){4}}
\put(13,1){\line(0,1){4}}
\put(3,3){\circle*{0.2}}
\put(5,3){\circle*{0.2}}
\put(11,3){\circle*{0.2}}
\put(5,1){\circle*{0.2}}
\put(13,1){\circle*{0.2}}
\put(13,3){\circle*{0.2}}
\put(7,3){\circle*{0.2}}
\put(15,3){\circle*{0.2}}
\put(5,5){\circle*{0.2}}
\put(13,5){\circle*{0.2}}
\put(2.5,3){\makebox(0,0){$i$}}
\put(7.5,3){\makebox(0,0){$j$}}
\put(5,0.5){\makebox(0,0){$s$}}
\put(5,5.5){\makebox(0,0){$r$}}
\put(10.5,3){\makebox(0,0){$r$}}
\put(15.5,3){\makebox(0,0){$s$}}
\put(13,0.5){\makebox(0,0){$i$}}
\put(13,5.5){\makebox(0,0){$j$}}
\end{picture}

{\footnotesize {\noindent\bf\footnotesize Figure 2:}
{\sl\footnotesize Geometric
relation between the pair of lattice sites
$<{\mbox{\boldmath $i$}}, {\mbox{\boldmath $j$}}>$ and the
pair of plaquette centers $<{\mbox{\boldmath $r$}} ,
{\mbox{\boldmath $s$}}>$.} }
\end{figure}
\noindent It is furthermore useful to define

$$\tau_{{\mbox{\boldmath $r$}}{\mbox{\boldmath $s$}}}^{x(y)}
= \cases{
1 &if $<{\mbox{\boldmath $i$}} , {\mbox{\boldmath $j$}}>$
is part of the loop (2.9b) (the loop (2.9c))
\cr 0 &otherwise \cr
} \eqno(2.13)$$
\noindent Henceforth we shall generally write
$\pi_{\mbox{\boldmath $rs$}}$ instead of
$\pi_{\mbox{\boldmath $ij$}}$ when  $< {\mbox{\boldmath $i$}}
, {\mbox{\boldmath $j$}}>$ and $<{\mbox{\boldmath $r$}}
, {\mbox{\boldmath $s$}}>$
are related as in Fig. 2. With this notation the result
of the
$\varphi_{{\mbox{\boldmath $ij$}}}$ integrations in
(2.12) is

$$Z_{M, N} = (2 \beta J)^{-MN} \sum_{n_x} \sum_{n_y}
{\sum_{\{
n_{{\mbox{\boldmath $r$}}} \}}}' \exp \left [ i
\sum_{<{\mbox{\boldmath $r$}} , {\mbox{\boldmath
$s$}}>} \pi_{{\mbox{\boldmath $r$}}
{\mbox{\boldmath $s$}}} \left ( n_{{\mbox{\boldmath
$r$}}} - n_{{\mbox{\boldmath $s$}}} + \sum_{\alpha = x,y}
\tau_{{\mbox{\boldmath $r$}} {\mbox{\boldmath
$s$}}}^{\alpha} \ n_{\alpha} \right ) \right ]$$
$$\times \ \exp \left [ - (4 \beta J)^{-1} \sum_{<
{\mbox{\boldmath $r$}} , {\mbox{\boldmath $s$}}>}
\left ( n_{{\mbox{\boldmath $r$}}} -
n_{{\mbox{\boldmath $s$}}} + \sum_{\alpha = x,y}
\tau_{{\mbox{\boldmath $r$}}
{\mbox{\boldmath $s$}}}^{\alpha} n_{\alpha}
\right )^2 \,\right ] \ \ \ . \eqno(2.14)$$

\noindent Eq. (2.14) represents the partition function
of a solid-on-solid model (or
``column model'') with Gaussian interaction, in which,
due to the randomness in the original
$XY$ model, some terms in occur with negative sign.
In the nonrandom
case with all
$\pi_{{\mbox{\boldmath $r$}} {\mbox{\boldmath $s$}}} = 0$,
we recover the partition function of the
well-known discrete Gaussian model, {\it summed} on different
step boundary conditions represented by the variables
$n_x$ and $n_y$.
\smallskip
\subsubsection{From SOS model to Coulomb gas}

The transformation from an SOS model to a Coulomb gas
is also
well-known [16]. We have to apply it here to the random
SOS model of
Eq. (2.14), taking properly into account again all
finite size effects.
For a function $f(n)$ of an integer variable $n$ one
has

$$\sum_{n = - \infty}^{\infty} f(n) = \sum_{q=-
\infty}^{\infty} \int_{- \infty}^{\infty} d \nu \
e^{2 \pi i q \nu} f(\nu) \ \ \ .  \eqno(2.15)$$
Applying this identity to the sum on the $\{
n_{{\mbox{\boldmath $r$}}} \}$ in (2.14) we
get

$$Z_{M,N} = (2 \beta J)^{-MN} \int_{- {1
\over 2}}^{{1 \over 2}} d \lambda \sum_{n_x} \sum_{n_y}
\sum_{\{q_{{\mbox{\boldmath $r$}}} \}} \int_{-
\infty}^{\infty} \prod_{{\mbox{\boldmath $r$}}} d
\nu_{{\mbox{\boldmath $r$}}} \times$$
$$\times \exp \left [ 2 \pi i \sum_{{\mbox{\boldmath
$r$}}} q_{{\mbox{\boldmath $r$}}}
\nu_{{\mbox{\boldmath $r$}}} + 2 \pi i \lambda \nu_0
+ i \sum_{<{\mbox{\boldmath $r$}} ,
{\mbox{\boldmath $s$}}>} \pi_{{\mbox{\boldmath $r$}}
{\mbox{\boldmath $s$}}} \left ( \nu_{{\mbox{\boldmath
$r$}}} - \nu_{{\mbox{\boldmath $s$}}} +
\sum_{\alpha} \tau_{{\mbox{\boldmath $r$}} {\mbox{\boldmath
$s$}}}^{\alpha} \ n_{\alpha} \right )- (4
\beta J)^{-1} \sum_{<{\mbox{\boldmath $r$}} ,
{\mbox{\boldmath $s$}}>} \left (
\nu_{{\mbox{\boldmath $r$}}} - \nu_{{\mbox{\boldmath
$s$}}} + \sum_{\alpha} \tau_{{\mbox{\boldmath $r$}}
{\mbox{\boldmath $s$}}}^{\alpha} \ n_{\alpha} \right
)^2 \right ] \eqno(2.16)$$

\noindent in which $\nu_0 \equiv
\nu_{{\mbox{\boldmath $r$}}_0}$ and the integral on
$\lambda$ takes care of the
condition $n_{{\mbox{\boldmath $r$}}_0} = 0$ in
(2.14). The integrations on the
$\nu_{{\mbox{\boldmath $r$}}}$ are Gaussian.
The most convenient way to carry them out requires
some preliminaries. We introduce the
coordinate representation ${\mbox{\boldmath $r$}}
= (x, y)$ with $x = 1, \cdots , M$ and $y = 1, \cdots , N$,
and the periodicity condition that $(x + M, y)$
and $(x, y + N)$ also denote the plaquette center
${\mbox{\boldmath $r$}}$. For horizontal and vertical
pairs $<{\mbox{\boldmath $r$}} , {\mbox{\boldmath
$s$}}>$ we shall have, by convention,
${\mbox{\boldmath $s$}} = (x + 1, y)$ and
${\mbox{\boldmath $s$}} = (x, y + 1)$, respectively.
For each
plaquette center ${\mbox{\boldmath $r$}}$ we define
a {\it frustration variable}
$p_{{\mbox{\boldmath $r$}}}$ by

$$2 \pi \ p_{{\mbox{\boldmath $r$}}} =
\widehat{\sum}^{{\mbox{\boldmath $r$}}}
\pi_{{\mbox{\boldmath $i$}} {\mbox{\boldmath $i'$}}} =
\pi_{01} - \pi_{20} - \pi_{30} + \pi_{04} \eqno(2.17)$$

\noindent where the notation is as in Eq. (2.8),
together with Fig. 3.
\begin{figure}
\setlength{\unitlength}{1cm}
\begin{picture}(19,6)
\thicklines
\put(7,3){\line(1,0){4}}
\put(9,1){\line(0,1){4}}
\put(9.2,2.7){\makebox(0,0){$0$}}
\put(6.5,3){\makebox(0,0){$3$}}
\put(11.5,3){\makebox(0,0){$1$}}
\put(9,0.5){\makebox(0,0){$4$}}
\put(9,5.5){\makebox(0,0){$2$}}
\put(8,2.7){\makebox(0,0){$\pi_{30}$}}
\put(8.7,2){\makebox(0,0){$\pi_{04}$}}
\put(10,2.7){\makebox(0,0){$\pi_{01}$}}
\put(8.7,4){\makebox(0,0){$\pi_{20}$}}
\put(8,3.3){\makebox(0,0){$-$}}
\put(9.3,2){\makebox(0,0){$+$}}
\put(10,3.3){\makebox(0,0){$+$}}
\put(9.3,4){\makebox(0,0){$-$}}
\put(8,3.3){\circle{0.5}}
\put(9.3,2){\circle{0.5}}
\put(10,3.3){\circle{0.5}}
\put(9.3,4){\circle{0.5}}
\put(7,3){\circle*{0.2}}
\put(9,3){\circle*{0.2}}
\put(11,3){\circle*{0.2}}
\put(9,1){\circle*{0.2}}
\put(9,5){\circle*{0.2}}
\end{picture}

{\footnotesize {\noindent\bf\footnotesize Figure 3:}
{\sl\footnotesize Diagram indicating the sign convention
in the definition (2.17) of the
frustration variable $p_{{\mbox{\boldmath $r$}}}$.}
The indices $0, 1, \cdots, 4$ are
shorthand for the plaquette centers ${\mbox{\boldmath
$r$}} , {\mbox{\boldmath $r$}}_1 , \cdots ,
{\mbox{\boldmath $r$}}_4$.}
\end{figure}
Let furthermore

$$\Pi_{\alpha} = \sum_{<{\mbox{\boldmath $r$}},
{\mbox{\boldmath $s$}}>}\!\!\! {}^{\alpha}\,\,
\pi_{{\mbox{\boldmath $r$}}
{\mbox{\boldmath $s$}}} \ \ \ ,  \eqno(2.18)$$

\noindent where the subscript $\alpha = x$ (or
$\alpha = y$) denotes restriction to vertical (or
horizontal) bonds (recall that a horizontal pair
$<{\mbox{\boldmath $i$}} , {\mbox{\boldmath $j$}}>$
corresponds to a vertical
pair $< {\mbox{\boldmath $r$}} , {\mbox{\boldmath $s$}}>$
and vice versa). At this point it is useful to fix the
positions of
the two loops around the torus by the specific choice

$$\tau_{{\mbox{\boldmath $r$}} {\mbox{\boldmath $s$}}}^x =
\delta_{y, N} \qquad \hbox{for} \ <{\mbox{\boldmath $r$}} ,
{\mbox{\boldmath $s$}}> = <(x , y),
(x, y + 1)>$$
$$\tau_{{\mbox{\boldmath $r$}} {\mbox{\boldmath $s$}}}^y =
\delta_{x, M} \qquad \hbox{for} \ <{\mbox{\boldmath $r$}}
, {\mbox{\boldmath $s$}}> = <(x , y),
(x + 1, y)> \eqno(2.19)$$
We pass in Eq. (2.16) from the $\nu_{{\mbox{\boldmath
$r$}}}$ to new variables of
integration $\nu '_{{\mbox{\boldmath $r$}}}$ defined by

$$\nu '_{{\mbox{\boldmath $r$}}} = \nu_{{\mbox{\boldmath
$r$}}} + M^{-1} x n_y + N^{-1} y n_x \eqno(2.20)$$

\noindent where as before ${\mbox{\boldmath $r$}}$ runs
through $(x, y)$ with $x = 1, \cdots , M$ and $y = 1,
\cdots , N$. After rearranging terms in the exponential
one finds for the partition function
(2.16)

$$Z_{M,N} = (2 \beta J)^{-MN} \int_{- {1 \over 2}}^{{1
\over 2}} d \lambda \sum_{n_x} \sum_{n_y}\exp \left [ -
2 \pi i \lambda (n_x + n_y) + i \left ( M^{-1} \Pi_y
n_y + N^{-1} \Pi_x n_x \right ) - (4 \beta J)^{-1} \left (
N^{-1} M n_x^2 + M^{-1} N n_y^2 \right ) \right ]$$
$$\times \sum_{\{q_{{\mbox{\boldmath $r$}}} \}} \exp
\left [ - 2 \pi i \sum_{{\mbox{\boldmath $r$}}}
q_{{\mbox{\boldmath $r$}}} \left (
M^{-1} xn_y + N^{-1} y n_x \right ) \right ]\times
Z'_{M,N} \eqno(2.21\mbox{a})$$

\noindent with
$$Z'_{M,N} = \int_{- \infty}^{\infty}
\prod_{{\mbox{\boldmath $r$}}} d
\nu '_{{\mbox{\boldmath $r$}}} \ \exp \left [ 2 \pi i
\sum_{{\mbox{\boldmath $r$}}} \nu
'_{{\mbox{\boldmath $r$}}} \left (
q_{{\mbox{\boldmath $r$}}} +
p_{{\mbox{\boldmath $r$}}} + \lambda
\delta_{{\mbox{\boldmath $r$}},
{\mbox{\boldmath $r$}}_0} \right )  -
(4 \beta J)^{-1} \sum_{<{\mbox{\boldmath $r$}} ,
{\mbox{\boldmath $s$}}>} (\nu '_{{\mbox{\boldmath $r$}}}
- \nu '_{{\mbox{\boldmath $s$}}} )^2
\right ] \ \ \ . \eqno(2.21\mbox{b})$$

\noindent We shall abbreviate

$$Q_{{\mbox{\boldmath $r$}}} = q_{{\mbox{\boldmath
$r$}}} + p_{{\mbox{\boldmath $r$}}} + \lambda
\delta_{{\mbox{\boldmath $r$}}, {\mbox{\boldmath
$r$}}_0} \ \ \ .
\eqno(2.22)$$

\noindent The integrals on the $\nu '_{{\mbox{\boldmath
$r$}}}$ in Eq. (2.21b) are now easily carried out with the
aid of the Fourier variables

$$\widehat{\nu}'_{{\mbox{\boldmath $k$}}} = (MN)^{- {1
\over 2}} \sum_{{\mbox{\boldmath $r$}}} e^{-i
{\mbox{\boldmath $k$}}.{\mbox{\boldmath $r$}}}
{\nu}'_{{\mbox{\boldmath $r$}}} \eqno(2.23)$$

\noindent and analogously defined
$\widehat{Q}_{{\mbox{\boldmath $k$}}}$, where

$${\mbox{\boldmath $k$}} = (k_x , k_y) = 2 \pi
\left ( M^{-1} \kappa_x , N^{-1} \kappa_y \right ) \ \ \ ,
\eqno(2.24\mbox{a})$$

$$\kappa_x = 0, 1, \cdots , M - 1 \ \ \ , \qquad
\kappa_y = 0, 1, \cdots , N-1 \ \ \ .
\eqno(2.24\mbox{b})$$
It is useful to make the specific choice
${\mbox{\boldmath $r$}}_0 = (M, N)$. Taking
properly care of the integration on
$\widehat{\nu}'_{\mbox{\boldmath $0$}}$, which is
exceptional and gives a factor $\delta
(\widehat{Q}_{\mbox{\boldmath $0$}})$, one
finds from (2.21b)

$$Z'_{M,N} = (MN)^{{1 \over 2}} \delta \left (
\sum_{{\mbox{\boldmath $r$}}} q_{{\mbox{\boldmath $r$}}}
+ \lambda \right ) C_{M,
N} \exp \left [ - 4 \pi^2 \beta J \sum_{{\mbox{\boldmath
$k$}} \not= {\mbox{\boldmath $0$}}}
\lambda_{{\mbox{\boldmath $k$}}}^{-1} |Q_{{\mbox{\boldmath
$k$}}}|^2 \right ] \eqno(2.25)$$

\noindent with

$$\lambda_{{\mbox{\boldmath $k$}}} = 4 \left ( \sin^2
{k_x \over 2} + \sin^2 {k_y \over
2} \right ) \ \ \ , \eqno(2.26)$$

$$C_{M,N} = \prod_{{\mbox{\boldmath $k$}} \not=
{\mbox{\boldmath $0$}}} \left ( 4 \pi \beta J /
\lambda_{{\mbox{\boldmath $k$}}} \right )^{{1 \over
2}} \ \ \ . \eqno(2.27)$$

\noindent Upon inserting (2.25) in (2.21a) one may carry
out the $\lambda$ integration with the
result
$$Z_{M,N} = (2 \beta J)^{-MN} (MN)^{{1 \over 2}} C_{M,N}
\sum_{n_x} \sum_{n_y}\exp \left [ i
\left ( M^{-1} \Pi_y n_y + N^{-1} \Pi_x n_x \right ) -
(4 \beta J)^{-1} \left ( N^{-1} M n_x^2 + M^{-1} N n_y^2
\right ) \right ]$$
$$\times \sum_{ \{ q_{{\mbox{\boldmath $r$}}} \}}{}'
\exp \left [ - 2 \pi i
\sum_{{\mbox{\boldmath $r$}}} \left ( M^{-1} x n_y +
N^{-1} y n_x \right )
- \beta {\cal H}_C \left ( \{ q_{{\mbox{\boldmath $r$}}} +
p_{{\mbox{\boldmath $r$}}} \} \right ) \right ] \ \ \ .
\eqno(2.28)$$

\noindent Here the prime restricts the summation to
neutral ``charge'' configurations $\{
{q}_{{\mbox{\boldmath $r$}}} \}$, i.e. satisfying

$$\sum_{{\mbox{\boldmath $r$}}} q_{{\mbox{\boldmath $r$}}}
= 0 \ \ \ , \eqno(2.29)$$

\noindent and ${\cal H}_C$ is the Coulomb Hamiltonian

$${\cal H}_C \left ( \left \{ q_{{\mbox{\boldmath $r$}}}
+ p_{{\mbox{\boldmath $r$}}} \right \} \right ) = 8 \pi^2 J
\sum_{{\mbox{\boldmath $r$}}} \sum_{{\mbox{\boldmath $r$}}\
'} U_{M,N} ({\mbox{\boldmath $r$}} -
{\mbox{\boldmath $r$}}\ ' )
(q_{{\mbox{\boldmath $r$}}} + p_{\mbox{\boldmath $r$}} )
(q_{{\mbox{\boldmath $r$}}'} + p_{{\mbox{\boldmath $r$}}'}
 ) \eqno(2.30)$$

\noindent where $U_{M,N}$ is the Coulomb potential on a
periodic lattice,

$$
U_{M,N}({\mbox{\boldmath $r$}}) = {1 \over 2MN}
\sum_{{\mbox{\boldmath $k$}}
\not={\mbox{\boldmath $0$}}} {e^{-i
{\mbox{\boldmath $k$}}.{\mbox{\boldmath $r$}}} - 1 \over
\lambda_{{\mbox{\boldmath $k$}}}} \ \ \ . \eqno(2.31)$$
This function has the periodicity properties

$$U_{M,N}({\mbox{\boldmath $r$}}) = U_{M,N} \left
( {\mbox{\boldmath $r$}} + M {\mbox{\boldmath $e$}}_1
\right ) = U_{M,N} \left ( {\mbox{\boldmath $r$}} +
N {\mbox{\boldmath $e$}}_2 \right ) \eqno(2.32)$$

\noindent where ${\mbox{\boldmath $e$}}_1 \equiv (1,
0)$ and ${\mbox{\boldmath $e$}}_2 \equiv (0,1)$. \par

The last step needed is to transform the sums on $n_x$
and $n_y$ in (2.28) according to Eq.
(2.15), by introducing continuous variables $\nu_x$ and
$\nu_y$, and new summation variables
$q_x$ and $q_y$. The integrations on $\nu_x$ and $\nu_y$
are again Gaussian and easily carried
out. The result can be slightly rewritten with the aid
of the relations

$$M^{-1} \left ( \Pi_y - 2 \pi \sum_{{\mbox{\boldmath $r$}}}
x q_{{\mbox{\boldmath $r$}}} \right ) = - 2 \pi M^{-1} P_x +
\pi_y \eqno(2.33\mbox{a})$$

$$N^{-1} \left ( \Pi_x - 2 \pi \sum_{{\mbox{\boldmath $r$}}}
y q_{{\mbox{\boldmath $r$}}} \right ) =
- 2 \pi N^{-1} P_y + \pi_x
\eqno(2.33\mbox{b})$$

\noindent where $\pi_y$ and $\pi_x$ are sums along loops
around the
torus,

$$\pi_y = \sum_{y=1}^N \pi_{(M, y), (1, y)} \ \ \ ,
\qquad \pi_x =
\sum_{x=1}^M \pi_{(x, N), (x, 1)}  \eqno(2.34)$$

\noindent and ${\bf P}$ is the electric dipole moment
$${\bf P} = (P_x , P_y) = \sum_{{\mbox{\boldmath $r$}}}
{\mbox{\boldmath $r$}} (q_{{\mbox{\boldmath $r$}}} + p_{{\bf
r}}) \ \ \ . \eqno(2.35)$$

\noindent The final result for $Z_{M,N}$ then becomes

$$Z_{M,N} = (2 \beta J)^{-MN} (MN)^{{1 \over 2}}
C_{M,N}\sum_{ \{q_{{\mbox{\boldmath $r$}}} \}}{}^{'}
\Theta_{N/M} \left ( N^{-1} P_y -
\pi_x/(2 \pi) \right ) \Theta_{M/N} \left ( M^{-1} P_x
- \pi_y/(2 \pi)
\right ) e^{- \beta {\cal H}_C} \eqno(2.36)$$

\noindent where

$$\Theta_a(u) = (4 \pi \beta Ja)^{{1 \over 2}} \sum_{q=-
\infty}^{\infty} \exp \left [ - 4 \pi^2
\beta Ja (q - u)^2 \right ] \eqno(2.37)$$

\noindent has the periodicity property $\Theta_a(u) =
\Theta_a(u+1)$, and
the prime refers to the charge neutrality condition
(2.29). \par

\subsection{Summary}

In view of the relatively technical character of the above
transformation, and for easy later reference, we summarize
here the
result. The partition function $Z_{M,N}$ (see Eq. (2.5))
of the $\pm J$
Villain XY model (Eqs. (2.3) and (2.4)) can be expressed
as the partition
function of a system of Coulomb charges
$\{q_{{\mbox{\boldmath $r$}}}\}$ on the dual
lattice, $$
Z_{M,N} = 2 \pi (2\beta J)^{-MN+1} (MN)^{1\over 2} C_{M,N}
\sum^{\infty}_{q_x, q_y = -\infty} \sum_{\{q_{{\mbox{\boldmath
$r$}}}\}}{}^{'} e^{-\beta
\hat {\cal H} (q_x, q_y, \{q_{{\mbox{\boldmath $r$}}}\})}
\eqno (2.38\mbox{a})
$$
where
$$
\hat {\cal H}(q_x, q_y, \{q_{{\mbox{\boldmath $r$}}}\}) =
4 \pi^2 J NM^{-1}\big [q_y + N^{-1}
\sum_{{\mbox{\boldmath $r$}}} y(q_{{\mbox{\boldmath $r$}}}
+ p_{{\mbox{\boldmath $r$}}}) +(2\pi)^{-1} \sum_{x=1}^M
\pi_{(x, N),(x, 1)}\big ]^2
$$
$$
+ 4 \pi^2 JN^{-1}M \big[ q_x + M^{-1}
\sum_{{\mbox{\boldmath $r$}}} x(q_{{\mbox{\boldmath $r$}}}
 +
p_{{\mbox{\boldmath $r$}}}) +(2\pi)^{-1} \sum_{y=1}^N
\pi_{(M,y), (1, y)}\big ]^2
$$
$$
+ 8 \pi^2 J \sum_{{\mbox{\boldmath $r$}}}
\sum_{{\mbox{\boldmath $r'$}}} U_{M,N} ({\mbox{\boldmath $r$}}
- {{\bf
r'}}) (q_{{\mbox{\boldmath $r$}}} + p_{{\mbox{\boldmath
$r$}}})(q_{{\mbox{\boldmath $r'$}}} + p_{{\mbox{\boldmath
$r'$}}}) \eqno
(2.38\mbox{b}) $$

The Coulomb potential $U_{M,N}$ and the constant $C_{M,N}$
are given by
(2.31) and (2.27), respectively, together with (2.26) and
(2.24). The
$q_{{\mbox{\boldmath $r$}}}$ run through all integer values
subject to the charge
neutrality condition (2.29). The sums on ${\mbox{\boldmath
$r$}}$ and ${{\mbox{\boldmath $r'$}}}$ run
through $(x, y)$ with $x = 1, \ldots, M$ and $y = 1,
\ldots, N$. Each
nearest neighbor bond $\langle {\mbox{\boldmath $r$}},
{\mbox{\boldmath $s$}} \rangle$ on the charge
lattice is dual to a nearest neighbor bond $\langle
{\mbox{\boldmath $i$}}, {\mbox{\boldmath $j$}}
\rangle$ of the original spin lattice, and the disorder
variables
$\pi_{{\mbox{\boldmath $rs$}}}$ that appear in (2.38)
are equal to the corresponding $\pi_{{\mbox{\boldmath
$ij$}}}$. The
frustration variables $p_{{\mbox{\boldmath $r$}}}$ are
defined in terms of the
$\pi_{{\mbox{\boldmath $rs$}}}$ by the diagram of
Fig. 3; they are half-integer
for a frustrated plaquette, and integer for an
unfrustrated plaquette. Even though the first two
terms in (2.38b) seem
to favor a definite coordinate representation of the
lattice, one may
verify with the aid of some algebra that $Z_{M,N}$ is
invariant under
translation of the origin, as of course it should be.

The transformation from the ferromagnetic 2D $XY$ model
to the Coulomb gas Hamiltonian ${\cal
H}_C$, including the charge neutrality condition, has
been known since
long [15]. Similarly
the replacement of the charges $q_{{\mbox{\boldmath $r$}}}$
by $q_{{\mbox{\boldmath $r$}}} + p_{{\mbox{\boldmath $r$}}}$
in the random case has
been known since Villain [1] and was treated in a more
general context
in Ref. [17]. Eqs. (2.38) show how on a {\it
finite lattice} the charge configurations receive a
supplementary weight,
represented by the first two terms in (2.38b), that depend
on the
electric dipole moment, in agreement with [18]. Whereas
this weight plays no r\^ole for a bulk
system in the thermodynamic limit, it is essential for
finite size
scaling considerations, and therefore (see Eqs. (1.2)
and (1.3)) for the
determination of the low temperature behavior of the
correlation
lengths. This fact will be exploited in the example
of Section 5 and in
the general heuristic theory of Section 6.

\section{Antiperiodic boundary conditions}

In view of the remark preceding Eq.
(2.3), it is very simple to change the periodic
boundary conditions in
the system discussed above into antiperiodic boundary
conditions (we
shall always take these along the seam joining the $M$th
and the 1st
column). It amounts to  changing $\pi_{{\mbox{\boldmath
$ij$}}}$ into
$\pi_{{\mbox{\boldmath $ij$}}} + \pi$ on the
antiferromagnetic seam. This just
means drawing another member of the  class of random
systems under
consideration. In particular, frustrated (unfrustrated)
plaquettes
remain frustrated (unfrustrated). The corresponding
antiperiodic
partition function $Z_{M,N}^{AP}$ differs from Eq. (2.36)
only in that
$\pi_x/(2 \pi)$ is replaced by $\pi_x/(2 \pi) + {1
\over 2}$.
This difference is at the origin of the finite size effect
of interest.
\par The following example, although trivial, shows how
ground state
energy differences are extracted from the final equations
of
Section 2. We compare the ground states of an $XY$
ferromagnet with
periodic and antiperiodic boundary conditions. The
energy difference
$\Delta E^{AP}$ is that of a spin wave of wavelength
$2M$, which can be
written down immediately :

$$\Delta E^{AP} = NMJ \left ( {2 \pi \over 2M} \right )^2
= \pi^2 {N
\over M} J \ \ \ . \eqno(3.1)$$
To see how the same result can be obtained from Eq. (2.36)
we use that
for both types of boundary conditions all $p_{{\mbox{\boldmath
$r$}}}$ vanish, and in
the ground state all $q_{{\mbox{\boldmath $r$}}}$ as well ;
furthermore that the
periodic boundary conditions have $\pi_x = 0$ and the
antiperiodic ones
$\pi_x = {1 \over 2}$. This gives

$$\Delta E^{AP} = - \lim_{\beta \to \infty} {1 \over \beta}
\log \Theta_{N/M}({1 \over 2})
\eqno(3.2)$$

\noindent which in view of (2.37) exactly coincides with
(3.1). \par

The $\Theta$ functions in Eq.(2.36), or equivalently, the
first two
terms in Eq.(2.38b), can be traced back mathematically to
the global
constraint on the spin variable differences
$\varphi_{{\mbox{\boldmath $ij$}}}$. This example
shows that they represent the energy of a continuous
spin wave
deformation forced into the system by the boundary
conditons. We shall
extend this interpretation to the case of general
disorder, and speak of
a ``{\em global spin wave}".\par

Ground state energy differences between P and AP boundary
conditions for more complicated
situations can also be obtained from (2.36), or (2.38),
at least in those cases where the ground state of the
charge system is known or can be plausibly guessed. A
nontrivial
example is discussed in Section 5.

\section{Reflecting boundary conditions}

Reflecting boundary conditions are introduced into
the $XY$ Hamiltonian by letting the spins
${\mbox{\boldmath $S$}}_{(1,n)}$ in the first lattice
column interact with the images of
the ${\mbox{\boldmath $S$}}_{(M,n)}$ in the last
column under reflection about an axis
in spin space. The counterpart of the periodic
Hamiltonian (2.3) is

$${\cal H}^R = J \sum_{<{\mbox{\boldmath $i$}},
{\mbox{\boldmath $j$}}>}\!\!\! {}^{reg}\,\, V \left (
\varphi_{{\mbox{\boldmath $i$}}} -
\varphi_{{\mbox{\boldmath $j$}}} - \pi_{{\mbox{\boldmath
$ij$}}} \right )
+ J \sum_{<{\mbox{\boldmath $i$}} , {\mbox{\boldmath
$j$}}>}\!\!\! {}^{exc}\,\, V \left (
\varphi_{{\mbox{\boldmath $i$}}} +
\varphi_{{\mbox{\boldmath $j$}}} - \pi_{{\mbox{\boldmath
$ij$}}} \right ) \ \ \ . \eqno(4.1)$$
Its partition function will be denoted by $Z_{M,N}^R$,
the upper index
$R$ in\-di\-ca\-ting, here and henceforth, reflecting
boundary
conditions. The subscript ``reg'' in (4.1) refers to
the regular terms
and the subscript ``exc'' to the exceptional ones,
modified by the
reflecting boundary conditions.
\subsection*{Transformation to a Coulomb gas}

The conversion of the Hamiltonian (4.1) into a Coulomb
gas Hamiltonian proceeds via the same
succession of transformations as in the case of periodic
boundary conditions. However, several
differences occur, and we shall indicate the main
modifications below. \par \smallskip
\hskip 1 truecm (i) When transforming to the variables
of Eq. (2.7), it is convenient to choose
${\mbox{\boldmath $i$}}_0 = (1, 1)$. The expressions
for $\varphi_{{\mbox{\boldmath $i$}}} +
\varphi_{{\mbox{\boldmath $j$}}}$ in the exceptional
terms in (4.1) then  become

$$\varphi_{(M,n)} + \varphi_{(1,n)} =
\varphi_{(M,n),(1,n)} - 2 \sum_{\ell = 2}^n \varphi_{(1,
\ell - 1), (1, \ell)} + 2 \varphi_0 \eqno(4.2)$$
$$\hskip 6 truecm (n = 1, 2, \ldots , N) \ \ \ .$$

\hskip 1 truecm (ii) The integration on $\varphi_0$
can be carried out only after the
introduction of the plaquette and loop variables
$\{ n_{{\mbox{\boldmath $r$}}} \}$, $n_x$, and $n_y$,
and leads
to the result $2 \pi \delta_{n_x,0}$. The integrations
on the $\varphi_{(M,n),(1,n)}$ and
$\varphi_{(1, n-1),(1,n)}$ are exceptional.
\smallskip

\hskip 1 truecm (iii) The discrete Gaussian partition
function, obtained as an intermediate
result, now becomes
$$Z_{M,N}^R = (2 \beta J)^{-MN} \sum_{n_y}
\sum_{\{n_{{\mbox{\boldmath $r$}}}\}}{}^{'}
\exp \left [ i \sum_{<{\mbox{\boldmath $r$}},
{\mbox{\boldmath $s$}}>}\! {}^{reg}\,
\pi_{{\mbox{\boldmath $r$}} {\mbox{\boldmath $s$}}}
(n_{{\mbox{\boldmath $r$}}} - n_{{\mbox{\boldmath $s$}}})
- (4 \beta J)^{-1}
\sum_{<{\mbox{\boldmath $r$}}, {\mbox{\boldmath $s$}}>}\!
{}^{reg}\, (n_{{\mbox{\boldmath $r$}}} -
n_{{\mbox{\boldmath $s$}}} )^2 \right
]$$  $$\times \exp \left [ i \sum_{<{\mbox{\boldmath
$r$}}, {\mbox{\boldmath $s$}}>}\! {}^{exc}\,  \pi_{
{\mbox{\boldmath $r$}}, {\mbox{\boldmath $s$}}} ( -
n_{{\mbox{\boldmath $r$}}} - n_{{\mbox{\boldmath
$s$}}} + n_y) - (4 \beta J)^{-1}
\sum_{<{\mbox{\boldmath $r$}}, {\mbox{\boldmath
$s$}}>}{}^{exc} (n_{{\mbox{\boldmath $r$}}} -
n_{{\mbox{\boldmath $s$}}} - n_y)^2
\right ]  \eqno(4.3)$$

\noindent in which, as before ${\mbox{\boldmath $r$}}
= (x, y)$ with $x = 1, \cdots , M$ and $y = 1, \cdots ,
N$, and the exceptional terms are the horizontal bonds
linking the $M$th to the first column ; we
have chosen $\tau_{{\mbox{\boldmath $r$}} {\mbox{\boldmath
$s$}}}^y = 1$ if $<{\mbox{\boldmath $r$}}, {\mbox{\boldmath
$s$}}>$ is exceptional (and
hence $\tau_{{\mbox{\boldmath $r$}} {\mbox{\boldmath
$s$}}}^y = 0$ if $<{\mbox{\boldmath $r$}} ,
{\mbox{\boldmath $s$}}>$ is regular) ; and the prime
restricts
the summation to configurations $\{n_{{\mbox{\boldmath
$r$}}}\}$ with $n_{(M,N)} = 0$. The exceptional terms
show
that between the columns $M$ and 1, in addition to
the step boundary condition represented by the
variable $n_y$, a reflection is imposed with respect
to the zero level of the column heights. \par
\smallskip

\hskip 1 truecm (iv) Upon continuing the succession of
transformations, one finds that
the integral on $\nu_y$ gives $2 \delta (2q_y +
\displaystyle{\sum_{{\mbox{\boldmath $r$}}}}
q_{{\mbox{\boldmath $r$}}} +
\lambda)$. After integration over $\lambda$ this
becomes $2 \delta
(2q_y - \displaystyle{\sum_{{\mbox{\boldmath $r$}}}}
q_{{\mbox{\boldmath $r$}}} , 0)$ where $\delta (a,
b) \equiv \delta_{a, b}$, and after summation on $q_y$
one gets $2
\delta ( \displaystyle{\sum_{\mbox{\boldmath $r$}}}
q_{\mbox{\boldmath $r$}} \ \mbox{mod} \ 2, 0)$, i.e.
the sum of the charges $q_{\mbox{\boldmath $r$}}$
should be even. \par \smallskip
\hskip 1 truecm (v) The relevant wavevectors are now

$${\mbox{\boldmath $k$}} = (k_x , k_y ) = 2 \pi \left
( M^{-1} \left ( \kappa_x + {1 \over 2} \right ), N^{-1}
\kappa_y \right ) \eqno(4.4)$$

\noindent with $\kappa_x$ and $\kappa_y$ as in (2.24b) ;
this set of
wavevectors will be
referred to by the upper index $R$ on summation and product
signs. We define in particular

$$C_{M,N}^R = \prod_{{\mbox{\boldmath $k$}}}{}^R \left (4
\pi \beta J/\lambda_{{\mbox{\boldmath $k$}}}
\right )^{{1 \over 2}}  \ \ \ . \eqno(4.5)$$
\smallskip
\hskip 1 truecm (vi) The variable $q_{{\mbox{\boldmath
$r$}}} - \delta_{x, M} \pi_{{\mbox{\boldmath $r$}}
{\mbox{\boldmath $s$}}}/\pi$, which
appears in the Coulomb Hamiltonian, may be renamed
$q_{{\mbox{\boldmath $r$}}}$ by a shift of the variables
$q_{(M,y)}$. \par \smallskip

The final result is

$$Z_{M,N}^R = 2(2 \beta J)^{-MN} C_{M,N}^R
\sum_{\{q_{{\mbox{\boldmath $r$}}}
\}}{}^{par}\, e^{- \beta {\cal H}_C^R} \eqno(4.6)$$

\noindent where the subscript ``par'' refers to the
parity condition

$$\left ( \sum_{{\mbox{\boldmath $r$}}}
q_{{\mbox{\boldmath $r$}}} + \pi_y/\pi \right )
\hbox{mod} \ 2
= 0 \ \ \ , \eqno(4.7)$$

\noindent and where

$${\cal H}_C^R \left ( \left \{ q_{\mbox{\boldmath $r$}}
+ p_{\mbox{\boldmath $r$}} \right \} \right ) = 8 \pi^2 J
\sum_{\mbox{\boldmath $r$}} \sum_{{\mbox{\boldmath $r$}}\
'} U_{M,N}^R
({\mbox{\boldmath $r$}} - {\mbox{\boldmath $r$}}\ ')
\left (
q_{\mbox{\boldmath $r$}} + p_{\mbox{\boldmath $r$}}
\right ) \left ( q_{{\mbox{\boldmath $r$}}\ '} +
p_{{\mbox{\boldmath $r$}}\ '} \right ) \eqno(4.8)$$

\noindent with ${\mbox{\boldmath $r$}}$ and
${\mbox{\boldmath $r$}}\ '$
in the range $(x, y), x = 1, \ldots , M, \ y = 1 ,
\ldots , N$, and with

$$U_{M,N}^R ({\mbox{\boldmath $r$}}) = {1 \over 2MN}
\sum_{{\mbox{\boldmath $k$}}}{}^R\,\, {e^{i
{\mbox{\boldmath $k$}}.
{\mbox{\boldmath $r$}}} \over \lambda_{{\mbox{\boldmath
$k$}}}} \ \ \ . \eqno(4.9)$$

This function has the (anti-)periodicity properties

$$U_{M,N}^R({\mbox{\boldmath $r$}}) = U_{M,N}^R \left
( {\mbox{\boldmath $r$}} + N {\mbox{\boldmath $e$}}_2
\right ) = - U_{M,N}^R \left (
{\mbox{\boldmath $r$}} + M {\mbox{\boldmath $e$}}_1
\right ) \ \ \ . \eqno(4.10)$$

\noindent Hence two charges at a fixed finite distance
on opposite sides of the re\-flec\-ting
boundary interact with each other's charge conjugated
image : reflecting boundary conditions
for the $XY$ spins lead to charge conjugating boundary
conditions for the Coulomb charges. \par

We now comment on this result and compare it to its
counterpart for
P and AP boundary conditions Eq. (2.36). The fact
that in the
partition function (4.6) no charge neutrality is
imposed becomes
understandable if one writes the interaction $U^R$ as

$$U_{M,N}^R({\mbox{\boldmath $r$}}) =
U_{2M,N}({\mbox{\boldmath $r$}}) - U_{2M,N}
({\mbox{\boldmath $r$}} + M {\mbox{\boldmath $e$}}_1
) \eqno(4.11)$$

\noindent which is easily checked. It is as though we
have a double system, of size $2M
\times N$ and with periodic boundary conditions, in
which every charge at a site ${\mbox{\boldmath $r$}}$
of the original system is paired up with an image charge,
equal but of opposite sign, at the
corresponding site ${\mbox{\boldmath $r$}} + M
{\mbox{\boldmath $e$}}_1$. Using Eq. (4.11) and
obvious symmetry properties
one can make this explicit by writing the Hamiltonian
(4.8) as

$${\cal H}_C^R \left ( \left \{ q_{\mbox{\boldmath $\rho$}} +
p_{\mbox{\boldmath $\rho$}} \right \}
\right ) = 4 \pi^2 J \sum_{\mbox{\boldmath $\rho$}}
\sum_{\mbox{\boldmath$\rho$}\ '} U_{2M,N}
(\mbox{\boldmath$\rho$} - \mbox{\boldmath$\rho$}\ ' )
(q_{\mbox{\boldmath $\rho$}} + p_{\mbox{\boldmath $\rho$}} )
(q_{\mbox{\boldmath $\rho$}\ '} + p_{\mbox{\boldmath
$\rho$}\ '})
 \ \ \ , \eqno(4.12)$$

\noindent where ${\mbox{\boldmath $\rho$}}$ and
${\mbox{\boldmath $\rho$}}\ '$ run through the $2 M
\times N$ lattice and it is understood that corresponding
sites carry
opposite charges. The extra prefactor ${1 \over 2}$ that
(4.12) has with
respect to (4.8) indicates that the energy density has
been smeared out
over twice the volume. The self-interaction present in
(4.8) has become
the interaction energy between a charge and its image
in (4.12).

\section{An example}

\subsection{Specific expressions for the disorder.
Ground state
and ground state energy with $P$ boundary conditions.}

As an example we consider an $M \times N$ lattice
containing a
rectangular array of $M' \times N'$ frustrated plaquettes,
consisting of the points
${\mbox{\boldmath $r$}}_{a, b} \equiv (x_a, y_b)$ with
$a = 1, 2, \ldots, M'$ and $b =
1, 2, \ldots, N' $, where $y_b = bN/N' \equiv b \ell_y$
with $\ell_y$
an even integer, and the $x_b$ are drawn randomly and
independently
with a density $M'/M \equiv 1/\ell_x$. The density of
frustrated
plaquettes is therefore $1/{\ell_x \ell_y}$. (Note
that drawing the frustrated
plaquettes randomly is not the same thing as drawing
the negative bonds
randomly [23, 24]; the latter procedure creates pairs
of frustrated
plaquettes, leading to a Coulomb gas with quenched
dipoles rather than
quenched charges). We let $Q^0_{{\mbox{\boldmath $r$}}}$
denote the ground state value
of $q_{{\mbox{\boldmath $r$}}} + p_{{\mbox{\boldmath
$r$}}}$. The arguments of this section rest on the
hypothesis that we can find the true ground state.
For $M'$ and $N'$
even we may safely assume that the ground state has
the chessboard array
of charges
$$ Q^0_{(x_a, y_b)} = \pm {1\over 2} (-1)^{a+b} \eqno
(5.1) $$
and zero charges on all the other plaquettes. The
ground state value of
${\cal H}_c$ therefore is, from (5.1) and (2.30),
$$
{\cal H}_c(\{ Q^0_{{\mbox{\boldmath $r$}}}\}) = 2 \pi^2
J \sum_{a,b} \sum_{a', b'}
U_{M,N} ({\mbox{\boldmath $r$}}_{a, b} - {\mbox{\boldmath
$r$}}_{a',b'})(-1)^{a+a'+b+b'} \eqno (5.2)
$$
Upon using the explicit form of the lattice Coulomb
potential, Eqs.
(2.31) and (2.26), we can rewrite (5.2) as a sum of
interaction
energies between the charge carrying lattice columns,
$$
{\cal H}_c(\{ Q^0_{{\mbox{\boldmath $r$}}}\}) = \pi^2
J \ell^{-2}_y N \sum_{a, a'}
(-1)^{a+a'} \sum_\nu V_\nu(x_a-x_a') \eqno (5.3)
$$
 where $\nu$ runs through the values $\pm {1\over 2}, \pm
{3\over 2},\ldots, \pm ({1\over 2} \ell_y -{1\over 2})$,
and in
which, with $c_\nu \equiv \cos 2\pi \nu \ell^{-1}_y$,
\setcounter{equation}{3}
\begin{equation}
\begin{array}{lcl}
V_\nu(x) &=& {1\over M} \sum\limits_{k_x} {e^{ik_xx}\over
4 - 2 \cos
k_x - 2c_\nu}\\[3mm]
{}&\simeq& {1\over 2} \big [ (2-c_\nu)^2-1\big ]^{-1/2}
\{ 2-c_\nu -
[(2-c_\nu)^2 - 1\big ]^{1/2}\}^{|x|}
\end{array}
\end{equation}
the last step involving the limit $M \rightarrow \infty$
at finite $|x|$. This
expression takes the much simpler form
$$
V_\nu(x) \sim (2 \pi |\nu|)^{-1} \ell_y \exp(-2 \pi|\nu x|
\ell_y^{-1}) \eqno (5.5)
$$
in the limit $\ell_x, \ell_y \rightarrow \infty$ with
$\ell_x/\ell_y$
fixed and $x$ of order $\ell_x$. In this limit, also,
the sum on
$\nu$ in (5.3) can be evaluated exactly and one finds
$$
{\cal H}_c (\{ Q^0_{{\mbox{\boldmath $r$}}}\}) \simeq -
\pi J \ell_y^{-1} N \sum_{a, a'}
(-1)^{a+a'} \log \hbox{tanh} (\pi |x_a -
x_{a'}|\ell_y^{-1}) \eqno
(5.6)
$$
This shows that the interaction energy is negative
between two
neighboring columns ($a$ and $a+1$), that it diverges
logarithmically at small distances $|x_a - x_{a'}|$,
as expected, and
tends to zero as $\exp(-\pi |x_a - x_{a'}| \ell^{-1}_y)$
for large $|x_a
- x_{a'}|$, so that it is short-ranged.

One remark is in place. The limit $\ell_x, \ell_y
\rightarrow \infty$
corresponds to a {\it dilute} array of frustrated
plaquettes. These must
be joined pairwise by long ladders of negative bonds
-- for which there
is no physical reason. The analysis that follows
will use Eq. (5.6);
however qualitatively similar results can be obtained
from Eqs. (5.3)
and (5.4), where $\ell_x$ and $\ell_y$ are arbitrary,
the key point
being that the large distance behavior of the
column-column interaction
is still determined by the $\nu = \pm {1\over 2}$
terms in Eq. (5.3). We
shall come back to the case of general $\ell_x$ and
$\ell_y$ in the discussion in Sec. 6.

\subsection{Comparison of $P$, $R$, and $AP$ boundary
conditions}

We now wish to consider the ground state energies and
their differences for the
three different boundary conditions, $P$, $AP$, and $R$.

\vspace{5mm}
(i) {\it Periodic boundary conditions}

For the frustrated plaquettes located on the array
defined in the
beginning of this section, we can always choose the
negative bonds such
that $\pi_x = \pi_y = 0$. Since the ground state
$\{Q^0_{{\mbox{\boldmath $r$}}}\}$ has
$P_x = P_y = 0$ (see Eq. (2.35)), the first two
terms in (2.38b) are
minimized by $q_x = q_y = 0$ and so do not contribute
to the ground
state energy, which is therefore entirely given by
(5.6).

One possible type of excitations from the ground
state are those
consisting of reversing all charges
$Q^0_{{\mbox{\boldmath $r$}}}$ in one column $a$.
The excitation energy is of the order of
$$
\Delta E(\xi_a) \equiv 2 \pi J\ell^{-1}_y N \log
\hbox{tanh}(\pi \xi_a
\ell^{-1}_y) \eqno (5.7)
$$
where $\xi_a$ is the distance between $x_a$ and the
closer one of
$x_{a-1}$ and $x_{a+1}$. Because of its exponential
decay with $|x_a -
x_{a'}|$, the interaction between $a$ and more distant
columns does
not change this estimate.

\vspace{5mm}
(ii) {\it Reflecting boundary conditions}

We first of all have to find the a priori unknown
ground state. To
this end we consider again the chessboard configuration
of charges of
Eq. (5.1) and calculate its energy ${\cal H}^R_C(\{
Q^0_{{\mbox{\boldmath $r$}}}\})$
under reflecting boundary conditions, starting from
Eqs. (4.8) and
(4.9). The result is again Eq. (5.6), but with
charge conjugating
boundary conditions, the interaction terms between
columns $a$ and
$a'$ on different sides of the reflecting boundary
have an extra minus
sign. Since the number $M'$ of  columns was supposed
even, this
means that there is somewhere a mismatch consisting
of two neighboring
columns $a$ and $a+1$ that are identically instead
of oppositely
charged. The energy cost is $\sim \Delta E(\xi)$,
with $\xi$ the
distance between these columns, and can be minimized
by shifting the
mismatch to the largest interneighbor distance. The
typical maximum
distance that can be found is
$$
\xi_{\hbox {max}} \simeq \ell_x \log \big ( {M\over
\ell_x}\big ) +
{\cal O}(1) \quad\qquad (M\to \infty) \eqno (5.8)
$$
and therefore the ground state energy difference
$\Delta E^R$ between
reflecting and periodic boundary conditions is of
the order of
$\Delta E (\xi_{\hbox{max}})$, which leads to
$$
\Delta E^R \sim \big ( {N\over \ell_y}\big ) \big
({M\over
\ell_x}\big)^{-\pi \ell_x/\ell_y}  \eqno (5.9)
$$

We can now link the finite size scaling of the ground
state to the low
temperature behavior of the correlation length as
indicated in the
introduction. Since the mechanism leading to Eq.(5.9)
is the reversal of
large domains of chiral variables, one deduces that
in case $\pi
\ell_x/\ell_y > 1$, there is in the thermodynamic limit
$M \sim N
\rightarrow \infty$ no chiral order at any finite
temperature $T$, and
that for $T\downarrow 0$ the chiral correlation length
$\xi_c$ diverges with
an exponent $\nu_c = 1/(1-\pi \ell_x/\ell_y)$. In case
$\pi \ell_x/\ell_y < 1$ (which
corresponds to relatively strong spatial anisotropy),
one deduces similarly that the system has a low
temperature phase with long range
chiral order. We shall come back to both cases
after discussing AP
boundary conditions.

\vspace{5mm}
(iii) {\it Antiperiodic boundary conditions}

Under AP boundary conditions the energy of a charge
configuration is
again given by (2.38b). In this case, ${\cal
H}_c(\{Q^0_{{\mbox{\boldmath $r$}}}\})$ is
as given by (5.6), but since now $\pi_y = 0$ and
$\pi_x = \pi$, the
energy of the charge configuration
$\{Q^0_{{\mbox{\boldmath $r$}}}\}$ has an extra
contribution $\pi^2 JNM^{-1}$ from the first two
terms in (2.38b),
which is exactly the spin wave energy encountered
in the discussion of
the  ferromagnet in Section 3. One cannot, however,
without further
inspection, identify this contribution as the ground
state
energy difference $\Delta E^{AP}$, since
$\{Q^0_{{\mbox{\boldmath $r$}}}\}$ is not
necessarily the ground state any more. The remainder
of the discussion
depends on the value of $\ell_x/\ell_y$. We discuss
first the case
$\pi \ell_x/\ell_y > 1$. Curiously, in this case
the mechanism to
construct a lower lying state is the same  as it was
for $R$
boundary conditions, viz. columnwise charge reversal.

Reversing a single column leaves $M^{-1} P_x = 0$ but
changes $N^{-1}
P_y$ from $0$ to ${1\over 2}$, which annihilates the
effect of $\pi_x
= \pi$ and therefore cancels the spin wave contribution
$\pi^2
JNM^{-1}$. The same cancellation can be obtained by
reversing the
charges in an odd number of otherwise arbitrary
columns. The energy
cost is the excitation energy of the columns. It
can be minimized by
choosing an odd number of consecutive columns $a, a+1,
\ldots a'$, such
that the distances $|x_{a-1} - x_a|$ and $|x_{a'+1} -
x_{a'}|$ are as
large as possible. Typically, they will again be of
order $\ell_x
\log(M/\ell_x)$. Hence we arrive for $\Delta E^{AP}$
at the estimate
$$
\Delta E^{AP} \sim \big ( {N\over \ell_y}\big )
\big ( {M\over
\ell_x}\big )^{-\pi\ell_x / \ell_y} \qquad \qquad
(\pi \ell_x/\ell_y >
1) \eqno (5.10)
$$
identical to $\Delta E^R$. This energy difference
is less than the
spin wave energy $\pi^2 JNM^{-1}$, and therefore
the ground state
will adjust to AP boundary conditions by a chiral
excitation, and not
by a spin wave excitation. In view of Eq.(1.3)
and our result for
$\nu_c$ found above we now conclude that $\nu_s
= \nu_c = 1/(1-\pi \ell_x/\ell_y)$. Hence at low
temperature the spin and chiral
correlation lengths, $\xi_s$ and $\xi_c$,  behave
as
$$
\xi_s \sim \xi_c \sim T^{-1/(1-\pi \ell_x/\ell_y)}
\qquad \qquad (\pi
\ell_x/\ell_y > 1) \eqno (5.11)
$$

In case $0 < \pi \ell_x/\ell_y < 1$ (large spatial
anisotropy, and,
as seen above, a low temperature phase with long
range chiral order),
the situation is different. The minimum excitation
energy of the
columns is still $(N/\ell_y)(M/\ell_x)^{-\pi
\ell_x/\ell_y}$, but
this is higher than the spin wave contribution
which it cancels.
Therefore in this case the ground state
$\{Q^0_{{\mbox{\boldmath $r$}}}\}$ remains
unchanged (and in the spin representation acquires
only an
additional global spin wave), and
$$
\Delta E^{AP} = \pi^2 J N M^{-1} \qquad \qquad (0 <
\pi
\ell_x/\ell_y < 1) \eqno (5.12)
$$
In the thermodynamic limit $M \sim N \rightarrow \infty$
it follows
(see Eqs. (1.2) and (1.3)) that $\nu_s = \infty$, which
is most
naturally (although not strictly necessarily) associated
with a low
temperature phase with power law decay of the spin-spin
correlation. In
this case the energy $\Delta E ^{AP}$ is not the result
of a domain
wall, but smeared out across the system.

All of the above discussion is for an even number
$M'$ of charge
carrying columns. We now comment on the differences
that intervene when $M'$ is odd. In that case the
value of $N^{-1} P_y$ in (2.36) is $\pm {1\over 4}$
and is changed to $\mp {1 \over 4}$ (modulo 1) under
replacement of P by AP boundary conditions. Hence
the value of $\Theta_{M/N}$ does not change and
(since $M^{-1} P_x =0$ both before and after) we
have $\Delta E^{AP} =0$.
The interpretation is that a global spin wave is
``caught'' in the
system and cannot release its energy. When we replace
P by R boundary
conditions, the global spin wave terms disappear from
the Hamiltonian
and release their energy so that $\Delta E^{R}$ is
negative. The
phenomenon of spin wave energy release was first
observed by Kawamura
and Tanemura [3] in their Monte Carlo simulations
of an $XY$ model with
general disorder. It appears here analytically,
but seemingly as a
consequence of the distinction between even and
odd $M'$ and hence closely tied to the present example.
But we shall see in the next section that it also
appears in the general case.

\section{General case}

On the basis of the experience gained we now attempt
a theory,
admittedly heuristic, for the general case. Let, as
before, ${\cal H}_{c}$
denote the standard lattice Coulomb Hamiltonian (2.30),
and ${\cal
H}_{c}^{R}$ its counterpart (4.8) with R boundary
conditions. Let in this
section ${\hat {\cal H}}^{P}$ denote the Hamiltinian
${\hat {\cal H}}$ of
Eq. (2.38b), and ${\hat {\cal H}}^{AP}$ its counterpart
with AP boundary
conditions. We shall take the unknown ground state
${\mbox{\boldmath{$Q$}}}^0$ of ${\cal H}_{c}$ as our
reference state and
denote its energy by $E_0$. Our strategy will be to try
to determine how
the ground states of ${\cal H}_{c}^{R}$ and ${\hat
{\cal H}}^{P,AP}$
differ from ${\mbox{\boldmath{$Q$}}}^0$.

Since the remaining discussion concerns the
low-temperature regime, it is convenient to work
with charges $\pm
{1 \over 2}$ on the frustrated plaquettes and charges
zero on all other plaquettes.
This simplification, that was also made by Villain [12],
leads to the
problem of an Ising model on a random lattice (namely,
the one composed
of centers of the frustrated plaquettes), with
logarithmically
increasing antiferromagnetic interactions and zero
magnetization.The logarithmic interaction will be
screened in a way that we cannot
precisely describe [it has to be if the energy per
charge is to be
finite] and it is reasonable to imagine that there
exist some effective
interaction decaying as a power law between spatially
separated neutral
sets of charges.

Assuming that we know the ground state
${\mbox{\boldmath{$Q$}}}^0$, the next
relevant question is what the lowest lying excitations
are. Clearly all
excitations can be described in terms of charge
reversals with respect
to the ground state ${\mbox{\boldmath{$Q$}}}^0$,
or alternatively, in terms
of contours on the dual lattice (given any reasonable
planar graph
representing the nearest neighbor relations on the
lattice of charges).
In the spirit of Fisher and Huse [26] we shall consider
the lowest lying
excitations involving the reversal of order $\sim L^2$
charges and
localized in an area of linear size $\sim 2L$ around a
preassigned point
in space. Such excitations will be called droplets.
Let their excitation
energy scale as $L^{-1/\nu_c}$, where, since we are
below $d_\ell$, the
exponent $\nu_c$ is positive. Due to the long-range
forces, the droplets
so defined need not constitute single domains.
[For short-range forces
they should and reduce to the droplets of Fisher
and Huse [26] theory.]
A typical scale-$L$ droplet may have to be
represented by a set of
disconnected contours. We shall assume, nevertheless,
that it is compact
enough so that there is always, typically, one main
contour enclosing
$\sim L^2$ charges. This amounts to assuming that a
domain wall can be
defined, whatever its width, around the reversed area.
Excitations that would not
fall in this category are, for example, those that
consist of many
small-size domains dispersed
in the volume
$L^2$, or fractal excitations extending throughout the
volume $L^2$ but
having fractal dimension less than $2$. Our hypothesis
is that such
reversals have excitation energies at least as high as
those of the
scale-$L$ droplets.

The ground states $\mbox{\boldmath $Q$}^{0,R}$ and
$\mbox{\boldmath
$Q$}^{0}$ of ${\cal H}_c^R$ and ${\cal H}_c$,
respectively, must differ
at least by a contour going around the torus in
the $y$ direction, that
is, by a scale-$N$ droplet. For the energy difference
between the two
ground states we therefore have
$$
E_0^R-E_0 \simeq c\, N^{-1/\nu_c} \eqno(6.1)
$$
where $c$ is a (positive or negative) random constant,
and $\nu_c$ is as
before the exponent of the chirality-chirality correlation
function, see
Eq. (1.2). This gives a way to determine this exponent.

We now turn to the boundary conditions P and AP. Let
$\mbox{\boldmath
$Q$}^{0,P}$ ($\mbox{\boldmath$Q$}^{0,AP}$) be the ground
state of ${\hat
{\cal H}}^P$ (${\hat{\cal H}}^{AP}$) and let $E_0^{P}$
($E_0^{AP}$) be
its energy. Since ${\hat{\cal H}}^{P,AP}$ differ from
${\cal H}_c$ by
the {\it addition} of two quadratic terms, we have
necessarily $E_0^{P,AP}
- E_0 \geq 0$. In the ground state $\mbox{\boldmath$Q$}^{0}$
these
additional terms take a random positive value of order $1$ (
at least, if we
assume that  $M^{-1} P_x$ and
$N^{-1} P_y$ do so in the ground state). This value can be reduced
by reversing
domains of charges
in $\mbox{\boldmath$Q$}^{0}$. The ground state
$\mbox{\boldmath$Q$}^{0,P}$ ($\mbox{\boldmath$Q$}^{0,AP}$)
is now determined by a compromise
between the minimization of ${\cal H}_c$ and the minimization
of the
additional terms, which drive the total electric dipole
moment to a
specific value. Since this is a global biasing force,
and since the
excitation energies go down with increasing length scale,
we expect that
the compromise leads to a $\mbox{\boldmath$Q$}^{0,P}$
($\mbox{\boldmath$Q$}^{0,AP}$) that differs
from $\mbox{\boldmath$Q$}^{0}$ by an excitation on the
scale $N$ of the
system. This excitation might not be of the same type
as the droplet
excitation discussed above, and therefore we put
$$
E_0^{P,AP} - E_0 \simeq c^{P,AP}\, N^{-1/\nu_s} \eqno(6.2)
$$
where $c^P$ and $c^{AP}$ are positive random constants
and the exponent
$\nu_s$ is also positive, it is the spin correlation
length exponent of
Eq. (1.2). But since, by hypothesis, the droplet excitations are
the lowest lying
ones that exist at each given scale, the excitation
that leads to (6.2)
is, at best, also a scale-$N$ droplet excitation in
which case $\nu_s =
\nu_c$, or is possibly a combination of droplet excitations
on smaller
scales in which case one might have $\nu_s^{-1} \leq
\nu_c^{-1}$. Hence
$$
\nu_s^{-1} < \nu_c^{-1}. \eqno(6.3)
$$
Upon combining these considerations we arrive at the
conclusion
$$\Delta E^{AP} \simeq (c^{AP}-c^P)\, N^{-1/\nu_s}$$
$$\Delta E^R \simeq -c^P\, N^{-1/\nu_s} - c\, N^{-1/\nu_c}
\eqno(6.4)$$
in which $c^{P,AP} \geq 0$ and $c$ may be of either
sign with
${\overline {c}}=0$. It follows that
${\overline{\Delta E ^R}} \simeq
-{\overline{c^P}} N^{-1/\nu_s}$, which is negative.
This is the
phenomenon of the release of spin wave energy as
one passes from P to R
boundary conditions. One sees furthermore that
$$
{\overline{(\Delta E ^{AP})^2}}^{1/2} \simeq
{\overline{(\Delta E
^{R})^2}}^{1/2} \sim N^{-1/\nu_s}
\eqno(6.5)
$$
i.e. both energy differences scale with the same
power of $N$.

We now discuss the two possibilities implied by
Eq. (6.3). If it holds
with the equality sign, we are in the case of a
single correlation
length for the spin and the chiral variables, as
found in the example
studied. This case needs no further comment. If
Eq. (6.3) were to hold
as a strict inequality, then there are two different
exponents $\nu_s$
and $\nu_c$. This would imply a longer correlation
length for the spin variables than for the chiral
variables. Moreover, $\Delta
E^R$ is then not the appropriate quantity to determine
$\nu_c$. In fact, in
this case $\nu_c$ cannot be found from a comparison of
P,AP and R
boundary conditions,
but should result from a comparison of ${\cal H}_c$
and ${\cal H}_c^R$ according to
Eq. (6.1).
Kawamura and Tanemura [3] extracted the exponent
$\nu_s$ from
a numerical determination of $\Delta
E^{AP}$ and $\nu_c$ from a postulated
expression involving both $\Delta E^{AP}$ and $
\Delta E^{R}$. In this
work there is a strict inequality that goes in the
sens opposite to (6.3).

The conclusion is that we find no
evidence for chiral order
extending on a longer length scale than spin order.

\section{Further comments and conclusions}

We have considered the two-dimensional $XY$ spin
glass with
$\pm J$ interactions. In sections 2 through 4 general
formulas are
presented that exhibit the interplay between chiral
and spin variables in
determining domain wall energies on finite $M\times
N$ lattices with
various types of boundary conditions. In
Section 5 we have considered, as an example, a specific
type of disorder
with infinite ranged correlations in the $y$ direction.
The spatial
distribution of the frustrated plaquettes depends on
two parameters
$\ell_x$ and $\ell_y$, whose ratio is an anisotropy
parameter. The
scaling properties of the domain wall energies with
system size lead us
to the following conclusions.

(i) {\it $XY$ spin glasses with critical temperature} $T=0$

For $\ell_x$, $\ell_y \rightarrow \infty$ and for
$\pi \ell_x/\ell_y
> 1$, the system of Sec. 5 has only a $T=0$ critical
point at which
the chiral and spin correlation lengths diverge with
the same
exponent $\nu_s = \nu_c = 1/(1-\pi\ell_x/\ell_y)$. We
add here
without proof that the phenomenon of a $T=0$ transition
with $\nu_s
= \nu_c$ holds in the entire region of the $\ell_x\ell_y$
plane
determined by
$$
4\ell_x(\ell_x - 1) \sin^2 {\pi\over 2\ell_y} > 1
\qquad \qquad
(\ell_x > 1 \qquad\hbox{and}\ \ell_y = 1, 2,
\ldots) \eqno (7.1)
$$
This phenomenon is analogous to what we found on
the one-dimensional
ladder [6] and tube [25] lattices. It is distinct,
however, from the
scenario proposed by Kawamura and Tanemura [2, 3]
and by Ray and
Moore [5] for the two-dimensional uncorrelated
random $\pm J$ $XY$
spin glass, according to which one would have $0 <
\nu^{-1}_c <
\nu^{-1}_s$. In order to rule out the
possibility that our
conclusions are restricted to a specific class of correlated
disorder, we consider in Section 6 the general case of
random disorder. The heuristic theory presented
 yields a different inequality, namely
$0 \leq \nu^{-1}_s \leq \nu^{-1}_c$. We therefore find
no evidence for chiral order extending on a longer scale
than spin order in two dimensions.

One may now go one
step further and speculate that since
 we had $\nu_s = \nu_c$ for $d=1$, the
mechanism
uncovered above, which enables the ground state to
accommodate to
a change from P to AP boundary conditions by a
low energy chiral
excitation, is general for uncorrelated random $\pm
J$ $XY$ model, and
that $\nu_s = \nu_c$ for all $d < d_\ell$.

(ii) {\it random $XY$ models with nonzero critical
temperature}

This class encompasses the model of Section 5 when
Eq. (7.1) is not
satisfied. In this case there is a low temperature
phase with long
range chiral order and, very probably, power law
decay of the
spin-spin correlations. This is the same scenario
that is believed
to hold for a fully frustrated two-dimensional $XY$
model [20-22],
which is, as a matter of fact, recovered in the
special limit
$\ell_x = \ell_y = 1$, and which is characterized
by $\nu_c^{-1} < 0
= \nu^{-1}_s$. It cannot, therefore, be identified
with the scenario
$\nu_c^{-1} < 0 < \nu_s^{-1}$ that Kawamura and
Tanemura [3] propose
for the {\it three}-dimensional random $\pm J$ $XY$
model.

Finally we
comment on the transition line $4\ell_x(\ell_x-1)
\sin^2(\pi/2\ell_y) = 1$ between the regions (i) and
(ii). Clearly
the distinction between these two types of ground state
behavior is
not due to the density of frustrated plaquettes, which
is $1/\ell_x
\ell_y$, but to the nature of correlations in their
spatial
arrangement. The different behaviors can be distinguished
by the
ground state response to a change between P and AP
boundary
conditions. In region (ii), the ground state is
elastic (it deforms
continuously), whereas in region (ii) it is ``brittle''
(it responds
by the formation of a chiral domain wall). This
transition appears in
this work somewhat as a byproduct of the analysis,
but it is
interesting in itself and merits a separate study.

\nocite{V,KT1,KT2,BD,RM,NHM,ON,NO,ON2,SY,K,V2,
MCMBC,K,JKKN,CW,FHS,MW,TJ,MS,LJNL,KU,HLDN,RSN,J,TNH}


\begin{thebibliography}{99}
\bibitem{V} J. Villain, J. Phys. C {\bf 10} (1977)
4793.
\bibitem {KT1} H. Kawamura and M. Tanemura, Phys. Rev.
B
{\bf 36} (1987) 7177.
\bibitem {KT2} H. Kawamura and M. Tanemura, J. Phys.
Soc. Japan {\bf 60}
(1991) 608.
\bibitem{BD} G.C. Batrouni and E. Dagotto, Phys.
Rev. B 37 (1988) 9875.
\bibitem {RM} P. Ray and M.A. Moore, Phys. Rev.
B {\bf 45}
(1992) 5361.
\bibitem{NHM} M. Ney-Nifle, H.J. Hilhorst, and
M.A. Moore,
Phys. Rev. B {\bf 48}(1993) 10254.
\bibitem {ON} Y. Ozeki and H. Nishimori, J. Phys.
Soc. Japan {\bf 57}
(1988) 4255.
\bibitem {NO} H. Nishimori and Y. Ozeki, J. Phys.
Soc. Japan {\bf 59}
(1990) 295.
\bibitem {ON2} Y. Ozeki and H. Nishimori, Phys.
Rev. B {\bf 46} (1992)
2879.
\bibitem{SY} M. Schwartz and A.P. Young, Europhys.
Lett.
{\bf 15} (1991) 209.
\bibitem {K} H. Kawamura, Phys. Rev. Lett. {\bf 68}
(1992) 3785.
\bibitem{V2} J. Villain, J. Physique {\bf 36} (1975)
581.
\bibitem {MCMBC}  B.W. Morris, S.G. Colborne, M.A.
Moore, A.J. Bray, and J.
Canisius, J. Phys. C {\bf 19} (1986) 1157.
\bibitem {K} H.J.F. Knops, Phys. Rev. Lett. {\bf 39}
(1977) 766.
\bibitem {JKKN} J.V. Jos\'e, L.P. Kadanoff, S.
Kirkpatrick, and D.R. Nelson,
Phys. Rev. B {\bf 16} (1977) 1217.
\bibitem {CW} S.T. Chui and J.D. Weeks, Phys.
Rev. B {\bf 14} (1976) 4978.
\bibitem {FHS} E. Fradkin, B.A. Huberman, and
S.H. Shenker, Phys. Rev. B
{\bf 18} (1978) 4789.
\bibitem {VB} A. Vallat and H. Beck, Phys. Rev.
B {\bf 50} (1994) 4015.
\bibitem {MW} B.M. McCoy and T.T. Wu, {\it The
Two-Dimensional Ising
Model} (Harward University Press, Cambridge, Mass., 1973).
\bibitem{TJ} S. Teitel and C. Jayaprakash, Phys.
Rev. B {\bf 27} (1983)
598.
\bibitem{KB} Y.M.M. Knops, B. Nienhuis, H.J.F.
Knops and H.W.J. Bl\"ote,
Phys. Rev. B {\bf 50} (1994) 1061.
\bibitem {RSJ} G. Ramirez-Santiago, J.V. Jos\'e,
Phys. Rev. B {\bf 49} (1994) 9567.
\bibitem{RSN} M. Rubinstein, B. Shraiman, and D.R.
Nelson, Phys. Rev. B
{\bf 27} (1983) 1800.
\bibitem {J} J.V. Jos\'e, Phys. Rev. B {\bf 20}
(1979) 2167.
\bibitem{TNH} M.J. Thill, M. Ney-Nifle, and H.J.
Hilhorst, in preparation.
\bibitem{FH} D.S Fisher and D.A. Huse,
Phys. Rev. B {\bf 38} (1988) 386.
\end{thebibliography}
\end{document}